\documentclass[runningheads]{llncs}

\usepackage[T1]{fontenc}
\def\doi#1{\href{https://doi.org/\detokenize{#1}}{\url{https://doi.org/\detokenize{#1}}}}
\usepackage{graphicx}
\usepackage{amsfonts}
\usepackage{algorithmic}
\usepackage{graphicx}
\usepackage{textcomp}
\usepackage{xcolor}
\usepackage{mathpartir}
\usepackage{mathtools}
\usepackage{todonotes}
\usepackage{enumitem}
\usepackage{ stmaryrd }

\DeclareMathOperator{\Assign}{\texttt{Assign}}
\DeclareMathOperator{\NDAssign}{\texttt{Obtn}}
\DeclareMathOperator{\Store}{\texttt{Store}}
\DeclareMathOperator{\From}{\texttt{Where}}
\DeclareMathOperator{\Seq}{\!\texttt{;}\ }

\DeclareMathOperator{\Jump}{\texttt{Jump}}
\DeclareMathOperator{\IJump}{\texttt{IndirJmp}}
\DeclareMathOperator{\Exit}{\texttt{Exit}}
\DeclareMathOperator{\Deref}{\texttt{!}\!}
\DeclareMathOperator{\Not}{\neg}
\DeclareMathOperator{\Separate}{\neq}
\DeclareMathOperator{\Aligned}{\equiv}
\DeclareMathOperator{\State}{\sigma}
\DeclareMathOperator{\op}{\oplus}
\DeclareMathOperator{\pre}{\tau}

\DeclareMathOperator{\blocks}{blocks}

\DeclareMathOperator{\evalP}{\longrightarrow_P}
\DeclareMathOperator{\evalB}{\longrightarrow_B}
\DeclareMathOperator{\evalS}{\longrightarrow_S}

\newcommand{\transition}[4]{#1~\colon~#2 #3 #4}
\newcommand{\evalE}[2]{#1 \vdash #2}

\DeclareMathOperator{\true}{\textit{True}}
\DeclareMathOperator{\false}{\textit{False}}

\newcommand{\jump}{\textsc{jump}}

\DeclareMathOperator{\memwrite}{\mathsf{write}}
\DeclareMathOperator{\memread}{\mathsf{read}}
\newcommand{\var}[1]{\ensuremath{\mathit{#1}}}

\newcommand{\nocompile}[1]{}
\newcommand{\nonterminal}[1]{\ensuremath{\underline{\mathsf{#1}}}}

\newcommand{\preP}[2]{\tau_P (#1,#2)}
\newcommand{\preB}[2]{\tau_B (#1,#2)}
\newcommand{\preS}[2]{\tau_S (#1,#2)}
\newcommand{\preStore}[3]{\tau_{\mathtt{store}} (#1,#2,#3)}
\newcommand{\register}[1]{\ensuremath{\mathtt{#1}}}

\usepackage{tikz}
\usetikzlibrary{shapes,calc,positioning,trees,arrows, automata, decorations.markings,decorations.pathmorphing}
\begin{document}
\title{Reachability Logic for Low-Level Programs}
\author{Nico Naus\inst{1}\and
Freek Verbeek\inst{1,2}\and
Marc Schoolderman\inst{3}\and
Binoy Ravindran\inst{1}}
\authorrunning{Naus et al.}
\institute{Virginia Tech, Blacksburg VA, USA \email{\{niconaus,freek,binoy\}@vt.edu}\and
Open Univeristy, The Netherlands \email{fvb@ou.nl}\and
Radboud University Nijmegen, The Netherlands \email{m.schoolderman@cs.ru.nl}}
\maketitle              % typeset the header of the contribution
\begin{abstract}
% !TEX root=../main.tex
%
%The abstract should briefly summarize the contents of the paper in
% Context
Automatic exploit generation is a relatively new area of research.
Work in this area aims to automate the manual and labor intensive task of finding exploits in software.
% Inquiry
In this paper we present a novel program logic to support automatic exploit generation.
% Approach
We develop a program logic called Reachability Logic, which formally defines the relation between reachability of an assertion and the preconditions which allow them to occur.
This relation is then used to calculate the search space of preconditions.
% Knowledge
We show that Reachability Logic is a powerful tool in automatically finding evidence that an assertion is reachable.
% Grounding
We verify that the system works for small litmus tests, as well as real-world algorithms.
An implementation has been developed, and the entire system is proven to be sound and complete in a theorem prover.
% Importance
This work represents an important step towards formally verified automatic exploit generation.
% Context: What is the broad context of the work? What is the importance of the general research area?
% Inquiry: What problem or question does the paper address? How has this problem or question been addressed by others (if at all)?
% Approach: What was done that unveiled new knowledge?
% Knowledge: What new facts were uncovered? If the research was not results oriented, what new capabilities are enabled by the work?
% Grounding: What argument, feasibility proof, artifacts, or results and evaluation support this work?
% Importance: Why does this work matter?

\end{abstract}
%
%
%

% !TEX root=../main.tex

\section{Introduction}

%CONTEXT/MOTIVATION: AEG
Exploit generation is the task of finding a security vulnerability in a program, as well as a way to leverage that vulnerability.
This is typically seen as a laborious manual task requiring intricate knowledge of the system under investigation.
In contrast, recent studies consider \emph{automated exploit generation} (AEG)~\cite{DBLP:conf/sp/BrumleyPSZ08,8432013}.
Even though AEG is ``still in its infancy''~\cite{DBLP:conf/ndss/AvgerinosCHB11}, the knowledge obtained by the exposed exploits has already proven to be valuable to guard and improve software.

Exploit generation, from an abstract point of view, is about reasoning over reachability of unwanted states.
For example, a common exploit is a buffer overflow that overwrites the return address, stored at the top of the stack frame.
For such exploits, the problem of AEG can be reformulated to deciding whether, just before a return, a state is reachable in which the return address has been modified with respect to the initial state.
Other types of exploits may aim to redirect control flow using indirect (dynamically computed) jumps.
For such exploits, the problem of AEG can be reformulated as a decision problem whether a state is reachable in which an indirection leads to an arbitrary address.
To summarize, a large part of AEG can be reduced to reasoning over reachability of assertions in low-level programs.

Current program logics are not suitable for reasoning over reachability of assertions.
Hoare Logic~\cite{DBLP:journals/cacm/Hoare69} reasons over all possible execution paths to prove correctness of a program with regards to some property.
In AEG, a vulnerable state will typically only occur in some corner case, making Hoare Logic unsuitable.
Reverse Hoare Logic~\cite{DBLP:conf/sefm/VriesK11} reasons over \textit{total reachability}.
This means that it defines the relation between a state and \textit{all} execution paths that lead to it.
Listing every possible path that leads to a certain state is difficult, if not unfeasible, and makes Reverse Hoare Logic unfeasible for AEG.
Knowing that there is one path that leads to the assertion under investigation is enough.

%CONTRIBUTION: JUMP
This paper proposes a novel theoretical foundation for reachability, that can be used for AEG.
The first step is to formalize an academic programming language similar to the well-known \textsc{While}~\cite{DBLP:journals/cacm/Hoare69} language.
Whereas \textsc{While} is intended to be an abstract model of high-level programming languages, this paper proposes \jump\ as an abstract model of low-level representations of executable behavior such as assembly or LLVM IR~\cite{DBLP:conf/cgo/LattnerA04}.
Reason is that AEG typically applies to executables: it considers low-level properties (e.g., stack overflows or return-oriented-programming) on low-level code.
The language \textsc{Jump} is characterized by being low-level, having unstructured control-flow (jumps instead of loops) and an unstructured flat memory model.
Moreover, it is \emph{non-deterministic}, allowing us to model the uncertainty of the semantics of various constructs found in executables~\cite{heule2016stratified,dasgupta2019complete}.

%CONTRIBUTION: LOGIC
In a similar fashion to Hoare Logic being defined for \textsc{While}, this paper proposes \emph{Reachability Logic} (RL) for \jump.
Reachability Logic consists of triples with a precondition, a program, and a postcondition.
Assuming the postcondition models a state, a reachability triple states that if the precondition holds, there exists at least one path to a state that satisfies the postcondition.
Whereas Hoare Logic is intended to reason over correctness, RL is thus intended to reason over \emph{reachability}.

%CONTRIBUTION: SEARCH SPACE
Hoare Logic comes with a proof methodology (precondition generation) that allows establishing correctness of a program.
In similar fashion, we provide a methodology for precondition generation for RL.
A key observation is that the nature of precondition generation between the two logics differ.
Hoare Logic requires over-approximative knowledge, e.g., loop invariants and full path exploration, as it is intended for verification.
Reachability Logic allows formal reasoning with under-approximative knowledge: no invariants or full path exploration is necessary.
We show that Reachability Logic defines a traversable \emph{search space}.
The precondition that shows reachability, if any paths exist, is somewhere in that search space.
The precondition generation is proven to be sound and complete; we know that the search space describes only actual reachability evidence, and that it describes all possible ways to reach the intended state.
The cost of having both soundness and completeness, is that the search space becomes infinite.
However, finding one path from assertion to initial state suffices and thus there is no need for full search space traversal to find exploits.

% EXPERIMENTAL RESULTS / PRACTICAL RESULTS
We demonstrate applicability of the proposed theory in two ways.
First, we implement a search space generator % and traverser
for RL and demonstrate on various litmus tests how a search space is generated, traversed and preconditions are found.
Second, we consider two realistic assembly implementations (Quicksort and Karatsuba multiplication) and model them in \textsc{Jump}.
We show how evidence of reachability is automatically generated.

To summarize, the contributions of this paper are:
\begin{itemize}
\item A novel logic for reasoning over reachability of assertions in low-level programs;
\item A formally proven correct precondition function that automatically computes initializations leading to states described by the postcondition;
\item The application of this logic to various litmus tests as well as two assembly implementations.
\end{itemize}

All results, source code and the formalized proof of correctness in Isabelle / HOL~\cite{paulson1994isabelle,kammuller1999locales,dawson2009isabelle} are publicly available~\footnote{https://github.com/niconaus/reachability-logic}.

% STRUCTURE
Section~\ref{sec:logic} introduces RL and argues why existing logics are not suitable for AEG.
Section~\ref{sec:exploit} and ~\ref{sec:precondition} introduce the \jump\ language and the exploit generation mechanism for it.
Litmus tests and applications are described in Sections~\ref{sec:lithmus} and~\ref{sec:casestudies}.
Section~\ref{sec:discussion} discusses the search space.
Related work is discussed in Section~\ref{sec:related} before we conclude in Section~\ref{sec:conclusion}.

% !TEX root=../main.tex

\section{Reachability Logic compared to other program logics}
\label{sec:logic}

\begin{table}[t]
\centering
\begin{tabular}{lcclr}\hline
\textbf{Name} & \multicolumn{3}{l}{\textbf{Definition}}  & \textbf{Usage}\\\hline
Hoare Logic & $\{P\} ~p~ \{Q\}$ & $\equiv$ & $\forall \State \cdot P(\State) \implies\forall \State' \cdot \State \xrightarrow{\ p\ } \State' \implies Q(\State')$ & Correctness\\
Reverse Hoare & $[P] ~p~ [Q]$     & $\equiv$ & $\forall \State' \cdot \, Q(\State') \implies \exists \State \cdot \State \xrightarrow{\ p\ } \State' \wedge~P(\State)$ & \hspace{-.4cm}Total reachability\\
Reachability Logic & $\langle P\rangle ~p~ \langle Q\rangle$ & $\equiv$ & $\forall \State \cdot P(\State) \implies \exists \State' \cdot \State \xrightarrow{\ p\ } \State' \wedge~Q(\State')$ & Reachability \\\hline\\\vspace{-.5cm}
\end{tabular}
\caption{Overview of program logics. Arrow $\xrightarrow{\ p\ }$ denotes the transition relation induced by program $p$.}
\label{tab:logic}
\vspace{-.8cm}
\end{table}

Reachability Logic (RL) revolves around triples $\langle P\rangle ~p~ \langle Q\rangle$, for some program $p$.
Here, $Q$ is a state-predicate that describes a state to be reached.
State-predicate~$P$ represents from where a $Q$ state is reachable: how must the machine be initialized for $Q$ to be possible?
These predicates thus formulate properties over states, which assign values to memory and registers, including heap, stack frame, and instruction pointer.

Table~\ref{tab:logic} provides definitions for both some existing logics and the proposed Reachability Logic.
Both Hoare Logic (HL) and Reverse Hoare Logic (RHL) can be extended with a frame rule, producing a ``separation'' version of the logic~\cite{DBLP:conf/cav/RaadBDDOV20}.
We consider the most elementary versions of these logics.
A reachability triple $\langle P\rangle ~p~ \langle Q\rangle$ expresses that for any state satisfying predicate $P$ there exists some non-deterministic execution such that a state satisfying $Q$ is reached.

\subsubsection*{Hoare Logic} HL~\cite{DBLP:journals/cacm/Hoare69} shows absence of bugs, with respect to some property, or, in other words, program \emph{correctness}.
We argue why it cannot be used for reachability.
%In the following arguments, let $Q$ denote a postcondition modeling a state to be reached.
First, consider an approach based on deriving an HL triple of the form $\{P\} ~p~ \{Q\}$.
This would express that all executions of the program lead to a desired state, which is unrealistic.
Typically only a few cases lead to the desired state and as such this approach would often lead to $P = \mathit{False}$ providing no information on reachability.
Second, consider an approach based on deriving a HL triple of the form $\{P\} ~p~ \{\neg Q\}$.
The intuition would be to derive preconditions $P$ that ensure that the state cannot be reached.
The negation of such a precondition may then be an initialization where the exploit does happen.

However, there are various counterarguments to this hypothetical approach.
First, by definition of Hoare triples, an initial state satisfying $\neg P$ does not necessarily lead to the Q-state, i.e, this approach does not show reachability.
Second, the negation of $P$ can grossly over-approximate the set of initial states that may lead to the Q-state, even in the case of a hypothetical perfect \emph{weakest} precondition generation (which is infeasible in practice due to scalability issues~\cite{barnett2005weakest}).

% Consider the following non-deterministic program $\mathtt{foo}$ and as exploit $Q = \mathtt{error} \equiv 1$.
% \begin{verbatim}
%   void foo(bool x, bool y) {
%     error = 1;
%     if (x && y && random()) error = 0;}
% \end{verbatim}
% The \emph{only} sound Hoare triple over this program would be:
% \[
%   \{ \mathit{False} \} ~\mathtt{foo(x,y)}~ \{ \neg Q \}
% \]
% Thus, this approach would consider any state (i.e., the negation of $\mathit{False}$) to be an exploit. However, in EL, the following triple holds:
% \[
%   \langle \mathtt{x} \wedge \mathtt{y} \rangle ~\mathtt{foo(i,v)}~ \langle Q \rangle
% \]
% This expresses that if $\mathtt{x}$ and $\mathtt{y}$ are initially true, the exploit is reachable, and there exists an execution of the program that leads from the initialization to the exploit.

As a final argument against the use of HL for reachability, we argue that HL inherently requires reasoning over \emph{all} execution paths.
It requires full path exploration and invariants over loops, which is known as the Achilles' heel of formal verification.
A logic to be used for AEG should not require any over-approximative reasoning: it reasons over \emph{some} path, or \emph{some} number of iterations of a loop.
A logic for AEG should not require loop invariants.

%If $\neg Q$ characterizes a bug, then an HL-triple ensures that for any possible execution starting in any $P$-state, the bug can never happen.
%HL requires the semantics of the program under investigation to be over-approxamitive.

% \begin{center}
% \begin{tabular}{cc}
% \begin{minipage}{0.5\textwidth}

% \end{minipage}
% \begin{minipage}{0.5\textwidth}
% \begin{verbatim}
% void bar(int64_t i, int64_t v) {
%   int64_t x[10];
%   x[(10 * 3^i) % 47] = v;
% }
% \end{verbatim}
% \end{minipage}
% \end{tabular}
% \end{center}

\subsubsection*{Reverse Hoare Logic} RHL~\cite{DBLP:conf/sefm/VriesK11} shows \emph{total reachability}.
Specifically, it shows that \emph{all} $Q$-states are reachable from a $P$-state.
RHL requires $P$ to characterize \emph{all} states leading to $Q$ which can easily become infeasible.
We illustrate this with the RHL triple below.
\[
  [w \equiv 100] ~v:=w~ [v>42]
\]
From precondition $w \equiv 100$, there is a valid path to a state satisfying postcondition $v>42$.
However, looking at the definition of RHL in Table~\ref{tab:logic}, this is \emph{not} a valid RHL triple.
For all states that satisfy the postcondition, there must be a state that satisfies the precondition that leads to it.
For example, $v=43$ is a state that satisfies the postcondition, but no state exists that satisfies $w \equiv 100$ leads to it.
RHL triples need to specify \emph{every state} that leads to the postcondition.
In the example above, this is trivial, but in more involved cases, this will quickly become infeasible.
In the case of loops, the precondition might be infinitely big, or require a manually written loop variant.
This makes RHL unsuitable as a foundation for AEG.
For the purpose of AEG, it satisfies to find one path, and indeed the following RL-triple holds: $\langle w \equiv 100 \rangle ~v:=w~ \langle v>42 \rangle$.
In other words, the universal quantification in RHL triples again requires a form of overapproximative reasoning (specifically: loop invariants), which should not be necessary in the case of AEG.

\subsubsection*{Incorrectness Logic} IL~\cite{DBLP:journals/pacmpl/OHearn20} has its basis in RHL, but with some changes.
The postcondition has an added exit-condition, indicating if the program terminated in an error or not. Like RHL, IL uses manually defined loop variants to deal with loops.
Finally, IL replaces conditionals by so called assumes, that assume the conditional to hold or not, depending on the selected branch.
While these changes make it easier to work with IL, the triples are still the same as in RHL.

%conclusion
Hoare Logic, Reverse Hoare Logic and Incorrectness Logic are not suitable for reasoning about reachability in the context of AEG.
The key difference between Reachability Logic and existing logics is that Reachability Logic is the only logic that satisfies \emph{all} of the following: 1.) RL is able to reason over the existence and reachability of paths, 2.) RL allows under-approximative semantics of the program $p$ under investigation, and 3.) is able to deal with non-determinism.
Reachability Logic instead does allow us to express the relation between a postcondition to be reached and an initial state configuration.
RL is suitable for not only describing this relation, but also automatically generating preconditions.
The coming sections describe RL in more detail, including precondition generation and examples to illustrate this process.

% Consider the program $\mathtt{bar}$ and some postcondition $Q \equiv x[10] = a$ below.
%
% \begin{verbatim}
% void bar(int64_t i, int64_t v) {
%   int64_t x[10];
%   x[(10 * 3^i) % 47] = v;
% }
% \end{verbatim}
%
% One might consider precondition $\mathtt{i} = 0 \wedge \mathtt{v} = a$, which indeed leads to an exploit.
% However, the following RHL triple does \emph{not} hold:
% \[
%   [\mathtt{i} = 0 \wedge \mathtt{v} = a] ~\mathtt{bar(i,v)}~ [Q]
% \]
% The exploit occurs as well when $i = 23$, since $10\cdot3^{23}$ is equal to $10$ (modulo $47$).
% RHL requires precondition $P$ to enumerate all values $i$ such that $10\cdot3^i = 10$, essentialy requiring to solve discrete logarithm.
%
% For AEG, it satisfies to find one exploit, and indeed the following EL-triple holds:
% \[
%   \langle \mathtt{i} = 0 \wedge \mathtt{v} = a \rangle ~\mathtt{bar(i,v)}~ \langle Q \rangle
% \]

% !TEX root=../main.tex

\section{The \jump\ language}\label{sec:exploit}

%Using the definition of exploit triples presented in the previous section, we can start formalizing the exploit search space.
%We build our formalization for a simple branching language we call \jump, inspired by the LLVM assembly language~\cite{lattner_adve_2010}.

%The exploit search space is defined in terms of precondition generation functions.
%These functions describe the relation between the exploit, the program statement and initial state.

The \jump\ language is intended as an abstract representation of low-level languages such as LLVM IR~\cite{DBLP:conf/cgo/LattnerA04} or assembly.
It has no explicit control-flow; instead it has jumps to addresses.
It consists of basic \emph{blocks} of elementary statements that end with either a jump or an exit.
Blocks are labeled with addresses.
Memory is modeled as a mapping from addresses to values (see Definition~\ref{def:mem} below).
Variables from a set $\mathcal{V}$ represent registers and flags.
The \emph{values} stored in variables or memory are words (bit-vectors) of type $\mathcal{W}$.

The following design decisions have been made regarding \jump.
\begin{description}
\item[Non-determinism]
We explicitly include \emph{non-determinism} through an $\NDAssign$ statement that allows to retrieve some value out of a set.
Non-determinism allows modeling of external functions whose behavior is unknown, allows dealing with uncertain semantics of assembly instructions and allows modeling user-input and IO.
The $\NDAssign$ statement is the only source of non-determinism in \jump.
\item[Unstructured memory]
Memory essentially consists of a flat mapping of addresses to values. There is no explicit notion of heap, stack frame, data section, or global variables. This is purposefully chosen as it allows to reason over pointer aliasing. For example, it allows Reachability Logic to formulate statements as ``the initial value of this pointer should be equal to the initial value of register \register{rsp}'' which is interpreted as a pointer pointing to the return address at the top of the stack frame.
\item[No structured control-flow]
All control flow happens through either conditional jumps or indirect jumps.
Indirect control flow is typically introduced by a compiler in case of switch-statements, callbacks, and to implement dynamic dispatch.
Note that a normal instruction such as the x86 instruction \texttt{ret} implicitly is an indirect jump as well.
\end{description}

%inspired by the LLVM assembly language.
%We have selected LLVM as a basis because many program languages can be compiled to it.
%As with LLVM, \jump\ has control flow in the form of jumps to code blocks.
%This means that there is no dynamic jumping.
%Control flow can only jump to a predefined number of code blocks.

\begin{definition}
A \jump\ \emph{program} $p$ is defined as the pair $(a_0,\blocks)$ where $a_0$ is the entry address, and $\blocks$ is a mapping from addresses to blocks.
A \emph{block} is defined by the grammar in Figure~\ref{fig:syntax}.
\end{definition}

\begin{figure}[t]\small
\begin{tabular}{l c l l}
  \multicolumn{4}{l}{Block}\\
   \nonterminal{b}     & = & $\nonterminal{s} \Seq \nonterminal{b} \mid  \Exit$ & Sequence, Exit\\
              &$\mid$& $\Jump \nonterminal{e}\ a_1\ a_2 \mid \IJump \nonterminal{e}$ & Conditional jump, Indirect jump\\
              &&&\\
  \multicolumn{4}{l}{Statement}\\
   \nonterminal{s} & = & $\Assign v\ \nonterminal{e}$ & Variable assignment\\
              &$\mid$ & $\NDAssign v\ \From~  \nonterminal{e}$ & Nondeterministic assign\\
              &$\mid$ & $\Store\ \nonterminal{e}\ v$ & Store v in address e\\
              &&&\\
\multicolumn{4}{l}{Expression}\\
 \nonterminal{e} & = & $w \mid v \mid \Deref \nonterminal{e}\mid \nonterminal{e_1}\ \op\ \nonterminal{e_2}\mid \Not \nonterminal{e}$ & Value, variable, deref, operation, negate\\
             %&$\mid$ &$v$ & Variable\\
             %&$\mid$ &$\Deref \nonterminal{e}$ & Dereference\\
             %&$\mid$ &$\Not \nonterminal{e}$ & Negation\\
             &&&\\
      $\op$  &$\in$& $\{+, -, \times, \%, <, \leq, \equiv, \neq, >, \geq, \land, \lor, \ldots \}$& Binary operators
\end{tabular}
\caption{\jump}
%\Description[definition of the \jump\ language]{definition of the \jump\ language}
\label{fig:syntax}
\vspace{-.4cm}
\end{figure}
A block consists of a sequence of zero or more statements, followed by either a conditional jump, indirect jump or $\Exit$.
The conditional jump jumps to the address $a_1$ only if the given expression evaluates to zero, otherwise to address $a_2$.
The indirect jump calculates the value of $\nonterminal{e}$ and jumps to the block at that address.
\emph{Statements} can be assignments or stores.
A deterministic assignment writes the value of expression $e$ to variable~$v$.
A nondeterministic assignment writes some value from a set to variable~$v$, defined by any value $w$ such that expression $e[v \coloneqq w]$ evaluates to non-zero.
Note that since expressions can read from memory, an assignment can model a load-instruction.
A store writes the value of variable $v$ into memory.
\emph{Expressions} consist of values, variables, dereferencing, binary operations and negation.

The state consists of values assigned to variables and memory.
We first define the memory model.

\begin{definition}\label{def:mem}
The tuple $(\mathcal{M}, \mathcal{R}, \memwrite, \memread)$ is a \emph{memory model}, where $\mathcal{M}$ is the type of the memory itself and $\mathcal{A}$ is the type of addresses.
Function $\memwrite$ is of type $\mathcal{A} \times \mathcal{W} \times \mathcal{M} \mapsto \mathcal{M}$ and function $\memread$ is of type $\mathcal{A} \times \mathcal{M} \mapsto \mathcal{W}$.
A memory model must satisfy:
\[
\begin{array}{l}
  \memread(a, \memwrite(a',w,m)) = \left\{
    \begin{array}{lcl}
      \memread(a,m) & \mbox{~if~} & a = a'\\
       w  & \mbox{~if~} & a \neq a'
    \end{array}\right.
\end{array}
\]
\end{definition}
%This memory model thus assumes that all accessed regions in memory are either aliasing or separate.
We assume values can bijectively be cast to addresses and we do so freely.
%Let $a$ and $a'$ be two values that represent addresses or pointers.
%We will use notation $a \Separate a'$ (resp. $a \Aligned a'$) to denote that these values correspond to separate (resp. aliasing) regions.

\begin{definition}
A \emph{state} $\sigma$ is a tuple $(\var{mem}, \var{vars})$ where $\var{mem}$ is of type $\mathcal{M}$ and $\var{vars}$ is of type $\mathcal{V} \mapsto \mathcal{W}$.
\end{definition}

Semantics are expressed through transition relations $\evalP$, $\evalB$ and $\evalS$ that respectively define state transitions induced by programs, blocks, and statements (see Figure~\ref{fig:semantics}).
For example, notation $\transition{p}{\State}{\evalP}{\State'}$ denotes a transition induced by program $p$ from state $\State$ to state $\State'$.
Notation $\evalE{\State}{e}=w$ denotes the evaluation of expression~$e$ in state $\State$ to value $w$.

\begin{figure*}[t]
  \begin{centering}
    \scriptsize
    $\inferrule*[left=prog]
        {\transition{\blocks(a_0)}{\State}{\evalB}{\State'}}
        {\transition{(a_0,\blocks)}{\State}{\evalP}{\State'}}$
    \quad
    $\inferrule*[left=seq]
        {{\transition{s}{\State}{\evalS}{\State'}}\\{\transition{b}{\State'}{\evalB}{\State''}}}
        {\transition{s \Seq\!\!\!b}{\State}{\evalB}{\State''}}$
    \quad
    $\inferrule*[left=jumpTrue]
        {{\evalE{\State}{e} \neq 0}\\{\transition{\blocks(a_1)}{\State}{\evalB}{\State'}}}
        {\transition{\Jump e\ a_1\ a_2}{\State}{\evalB}{\State'}}$
    \\[1.5ex]
    $\inferrule*[left=jumpFalse]
        {{\evalE{\State}{e} = 0}\\ {\transition{\blocks(a_2)}{\State}{\evalB}{\State'}}}
        {\transition{\Jump e\ a_1\ a_2}{\State}{\evalB}{\State'}}$
  \quad%[1.5ex]
  $\inferrule*[left=indirectJump]
      {{\evalE{\State}{e} = a}\\ {\transition{\blocks(a)}{\State}{\evalB}{\State'}}}
      {\transition{\IJump e}{\State}{\evalB}{\State'}}$
      \quad
      $\inferrule*[left=exit]
               { }
               {\transition{\Exit}{\State}{\evalB}{\State}}$
    \\[1.5ex]
    $\inferrule*[left=assign]
        {\evalE{\State}{e} = w}
        {\transition{\Assign v\ e}{(mem,vars)}{\evalS}{~(mem,vars[v \coloneqq w])}}$
    \quad
    $\inferrule*[left=store]
        {\evalE{(mem,vars)}{e_1} = a \\ \evalE{\State}{e_2} = w}
        {\transition{\Store e_1\ e_2}{(mem,vars)}{\evalS}{~(\memwrite(a,w,mem),vars)}}$
    \\[1.5ex]
    $\inferrule*[left=ndassign]
        {\evalE{\State}{e[v \coloneqq w] \neq 0}}
        {\transition{\NDAssign v\ \From~ e }{(mem,vars)}{\evalS}{~(mem,vars[v \coloneqq w])}}$
    \quad%\[1.5ex]
    $\inferrule*[left=load]
        {{\evalE{(mem,vars)}{e} = a}\\{\memread(a,mem) = w}}
        {\evalE{(mem,vars)}{\Deref e} = w}$\\
  \end{centering}
  \caption{Semantics of \jump. Rules for evaluation of expressions are omitted, except for the dereference operator.}
%\Description[Semantics of \jump]{Semantics of \jump}
  \label{fig:semantics}
  \vspace{-.4cm}
\end{figure*}

The semantics are largely straightforward.
A program is evaluated by evaluating the block pointed to by the entry address.
A conditional jump is evaluated by evaluating the condition, and then the target block.
Indirect jumps are evaluated in a similar manner, by evaluating the expression to obtain the block to jump to.
Non-standard is the nondeterministic assignment $\textsc{ndassign}$, which evaluates expression $e$ after substituting the variable $\var{v}$ for some value $w$.
For any value $w$ where expression $e$ evaluates to non-zero, a transition may occur.
A \textsc{store} evaluates expression $e$ producing some address $a$, and writes the value of variable $v$ to the corresponding region in memory.
A \textsc{load} uses function $\memread$ to read from memory.

\section{Precondition generation}\label{sec:precondition}

This section formalizes precondition generation functions for Reachability Logic.
The central idea is to formulate a transformation function $\tau$ that takes as input 1.) a program $p$, and 2.) a postcondition $Q$, and produce as output a disjunctive \emph{set} of preconditions.
This transformation function follows the recursive structure of \jump, i.e., we formulate functions $\tau_P$, $\tau_B$ and $\tau_S$ that perform transformations relative to a program, a block and a statement respectively.

When applied statement-by-statement, these functions populate the Reachability search space.
This search space is an acyclic graph, with symbolic predicates as vertices and as root the initial postcondition.
It contains a labeled edge $(Q,s,P)$ if and only if application of function $\tau_S$ for statement~$s$ and postcondition~$Q$ produces a set containing precondition~$P$.

%Figure~\ref{fig:pred} lists the definition of symbolic predicates that form pre- and postconditions.

Below, the definition of predicates that form pre- and postconditions is given.

\begin{table}
  \begin{center}
  \begin{tabular}{l c l l}
    \multicolumn{4}{l}{Predicate}\\
    \quad\nonterminal{P} & = & $\exists\ i \in \nonterminal{e}~\cdot~\nonterminal{P} \mid \nonterminal{e}$ & \quad Existential quantification, expression\\
  %                 &$\mid$ & $\nonterminal{e_1} ~\Separate~ \nonterminal{e_2}$& Separate region condition\\
  %                 &$\mid$ & $\nonterminal{e_1} ~\Aligned~ \nonterminal{e_2}$& Aliasing region condition\\
  \end{tabular}
    \end{center}
  \vspace{-.8cm}
\end{table}

Predicates $\nonterminal{P}$ are expressions (true if and only if they evaluate to non-zero), but can also contain outermost existential quantifiers.
The predicate $\exists i \in e \cdot P$ means there exists a value $w$ for $i$ such that both $e[i \coloneqq w]$ and $P[i \coloneqq w]$ hold.
%Moreover, predicates can explicitly denote that the values represented by two expressions refer to separated or aliasing memory regions.

% \begin{figure}
%   \begin{center}
%     \begin{tabular}{l c l l}
%       \multicolumn{4}{l}{Predicate}\\
%       \quad\nonterminal{P} & = & $\exists\ i \in \nonterminal{e}~\cdot~\nonterminal{P}$ & Existential quantification\\
% %                 &$\mid$ & $\nonterminal{e_1} ~\Separate~ \nonterminal{e_2}$& Separate region condition\\
% %                 &$\mid$ & $\nonterminal{e_1} ~\Aligned~ \nonterminal{e_2}$& Aliasing region condition\\
%                  &$\mid$ &$\nonterminal{e}$& Expression
%     \end{tabular}
%   \end{center}
%
%   \caption{Predicate definition}
% %  \Description[Predicate definition]{Predicate definition}
%   \label{fig:pred}
% \end{figure}

Given a program $p$ and a postcondition $Q$ defined in the predicate language above, a transformation is \emph{sound} if it generates preconditions $P$ that form a reachability triple.
Soundness means that a generated precondition actually represents an initial state that non-deterministically leads to the Q-state.
To define soundness, we first define the notion of a reachability triple relative to blocks (instead of a program as a whole as in Section~\ref{sec:logic}):
\begin{definition}
A reachability triple for block $b$ is defined as:
\[
  \langle P\rangle ~b~ \langle Q\rangle \equiv \forall \State \cdot P(\State) \implies \exists \State' \cdot~Q(\State') \wedge (\transition{b}{\State}{\evalB}{\State'})
\]
\end{definition}
We repeat this definition to stress that a reachability triple over block $b$ intuitively means that precondition $P$ leads to the desired state when running the block \emph{and subsequent blocks jumped to, until an exit}, i.e, not just running the instructions within block~$b$ itself.
This is due to the nature of transition relation $\evalB$ (see Figure~\ref{fig:semantics}).
A similar definition can also be made for statements: a reachability triple $\langle P\rangle ~s~ \langle Q\rangle$ for statement $s$ is defined for transition relation $\evalS$ and thus concerns the execution of the individual statement $s$ only.

\begin{definition}
Function $\tau_P$ is \emph{sound}, if and only if, for any program $p$ and postcondition $Q$:
\[
  \forall P\in \preP{p}{Q} \cdot \langle P\rangle ~p~ \langle Q \rangle
\]
\end{definition}
Similarly, soundness is defined for blocks and statements.

Figure~\ref{fig:preP} shows the transformation functions.
Function $\pre_P$ starts at the entry block of the program.
The program is then traversed in the style of a \emph{right fold}~\cite{sheard1993fold}: starting at the entry the program is traversed up to an exit point, from which postcondition transformation happens.
Function $\pre_B$ is identical to standard weakest-precondition generation in the cases of sequence and exit.
In the case of a conditional jump, two paths are explored.
Both could lead to exploits, as long as the branching conditions remain internally consistent.
In case of an indirect jump, all possible addresses that can be jumped to, are explored.

\begin{figure}[t]
  \begin{center}\scriptsize
\begin{tabular}{@{}l c l}
  \multicolumn{3}{@{}l}{Program:}\\
  $\preP {p} {Q}$  & $=$ & $\preB{\blocks(a_0)}{Q}$\\[1ex]

  \multicolumn{3}{@{}l}{Block:}\\
  $\preB{\nonterminal{s} \Seq\!\!\!\nonterminal{b}}{Q}$ &$=$& $\bigcup \{ \preS {\nonterminal{s}}{P} \mid P \in \preB{\nonterminal{b}}{Q} \}$\\
  $\preB{\Jump \nonterminal{e}\ a_1\ a_2}{Q} $ &$=$ &$\{P_1\land \nonterminal{e} \mid P_1 \in \pre_b(\blocks(a_1),Q)\}\cup\{ P_2 \land \neg \nonterminal{e}\mid P_2 \in \pre_b(\blocks(a_2),Q)\}$\\
  $\preB{\IJump \nonterminal{e}}{Q} $ &$=$ & $\{P\land \nonterminal{e} \equiv a \mid P\in  \pre_b(\blocks(a),Q), a \in \operatorname{dom}(\blocks)\}$\\
  $\preB{\Exit}{Q}$ & $=$ & $\{Q\}$\\[1ex]

  \multicolumn{3}{@{}l}{Statement:}\\
  $\preS{\Assign\ v\ \nonterminal{e}}{Q}$    & $=$ & $\{Q[v\coloneqq \nonterminal{e}]\}$ \\
  $\preS{\NDAssign\ v\ \From\nonterminal{e}}{Q}$ & $=$ & $\{\exists i \in \nonterminal{e} \cdot Q[v \coloneqq i]\}$\\
  $\preS{\Store\ \nonterminal{e}\ v}{Q}$     & $=$ & $\{Q' \land P \mid (Q',P) \in\preStore{\nonterminal{e}}{v}{Q}\} $
\end{tabular}
\end{center}
\caption{Transformation functions for a program $p = (a_0,\blocks)$.}
%\Description[transform function for predicates]{transform function for predicates}
\label{fig:preP}
\end{figure}

Function $\pre_S$ is standard in case of deterministic assignment.
In case of nondeterministic assignment, according to the execution semantics, some value $i$ needs to be found that fulfills the condition $\nonterminal{e}$.
That existentially quantified value is substituted for variable $v$ in the postcondition.

In the case of memory assignment, predicate transformation is a bit more complex.
%It needs to generate conditions over memory, in terms of separation and aliasing relations, under which the exploit is possible.
Consider the following example:
\begin{align*}
  \Store\ x\ 42\Seq \Store\ y\ 43\ \langle\Deref x \equiv 42\rangle
\end{align*}
If $x$ and $y$ alias, then $!x$ will be 43 after execution.
The postcondition $\Deref x \equiv 42$ can only hold if~$x$ and~$y$ are separate.

We explicitly encode assumptions about memory separation into the generated preconditions.
The $\pre_{\mathtt{store}}$ function listed in Figure~\ref{fig:storeCases} takes care of this.
It takes as input the address~$a$ to which value $v$ is written, and the postcondition $P$.
It returns a set of tuples $(Q,Q_{mem})$ where $Q$ is the precondition and $Q_{mem}$ provides the pointer-relations under which that substitution holds.
For example, we have $\preStore{a_1}{v}{\Deref a_2\equiv 42} = \{(\Deref a_2\equiv 42, a_1 \Separate a_2),(v\equiv 42,a_1 \Aligned a_2)\}$. This indicates two possible substitutions when transforming postcondition into precondition:
\[
\begin{array}{ccccl}
  \langle \Deref a_2 \equiv 42 \rangle &\Store\ a_1\ v &\langle \Deref a_2 \equiv 42 \rangle & \mbox{~if~} & a_1 \Separate a_2 \\
  \langle v \equiv 42 \rangle &\Store\ a_1\ v& \langle \Deref a_2 \equiv 42 \rangle & \mbox{~if~} & a_1 \Aligned a_2
\end{array}
\]
All other cases of $\pre_{\mathtt{store}}$ merely propagate the case generation.

%It adheres to the following specification.

%\begin{definition}
%Function $\tau_{\mathtt{store}}$ is sound if and only if, for any epxression $e$ and postcondition $Q$:
%\[
%  \State=(var,mem),\evalE {\nonterminal{e}} \State = v \cdot \preStore{e}{v}{Q}(\State) \iff Q(\State')
%\]
%\end{definition}

\begin{figure}[t]
  \begin{center}\scriptsize
\begin{tabular}{@{}l c l}
  $\preStore{a}{v}{c}$ & $=$ & $\{(c,\true)\}$\\
  $\preStore{a}{v}{x}$ & $=$ & $\{(x,\true)\}$\\
  $\preStore{a_1}{v}{\Deref a_2}$ & $=$& $\{(\Deref a_2, a_1 \Separate a_2),(v,a_1 \Aligned a_2)\}$\\
  $\preStore{a}{v}{\Not \nonterminal{e}}$ &$=$& $\{(\Not \nonterminal{e}',p)\mid (\nonterminal{e}',p) \in \preStore{a}{v}{\nonterminal{e}}\}$\\
  $\preStore{a}{v}{\nonterminal{e}_1 \op \nonterminal{e}_2}$ & $=$& $\{(\nonterminal{e}_1' \op \nonterminal{e}_2',p_1\land p_2)\mid (\nonterminal{e}_1',p_1) \in \preStore{a}{v}{\nonterminal{e}_1}
  ,(\nonterminal{e}_2',p_2)\in \preStore{a}{v}{\nonterminal{e}_2}\}$\\
%  $\preStore{r}{v}{p_1\land p_2}$ &$=$& $\{(p_1'\land p_2',p_3\land p_4)\mid (p_1',p_3) \in \preStore{r}{v}{p_1}, (p_2',p_4)\in \preStore{r}{v}{p_2}\}$\\
  $\preStore{a}{v}{\exists i \in\ \nonterminal{e} \cdot \nonterminal{P}}$&$=$& $\{(\exists i \in \nonterminal{e}' \cdot \nonterminal{P}',p_1 \land p_2)\mid(\nonterminal{e}',p_1) \in \preStore{a}{v}{\nonterminal{e}}
  , (\nonterminal{P}',p_2) \in \preStore{a}{v}{\nonterminal{P}}\}$
%  $\preStore{r}{v}{\nonterminal{e}_1 \Separate \nonterminal{e}_2}$ &$=$& $\{(\nonterminal{e}_1' \Separate \nonterminal{e}_2',p_1\land p_2)\mid (\nonterminal{e}_1',p_1) \in \preStore{r}{v}{\nonterminal{e}_1}\},(\nonterminal{e}_2',p_2) \in \preStore{r}{v}{\nonterminal{e}_2}\}$\\
%  $\preStore{r}{v}{\nonterminal{e}_1 \Aligned \nonterminal{e}_2}$ &$=$& $\{(\nonterminal{e}_1' \Aligned \nonterminal{e}_2',p_1\land p_2)\mid (\nonterminal{e}_1',p_1) \in \preStore{r}{v}{\nonterminal{e}_1},(\nonterminal{e}_2',p_2) \in \preStore{r}{v}{\nonterminal{e}_2}\}$
\end{tabular}
\end{center}
\caption{Case definitions for precondition of store}
%\Description[cases for precondition of store]{cases of precondition for store}
\label{fig:storeCases}
\vspace{-.4cm}
\end{figure}

There are no special rules for dealing with loops.
Instead, loops are unrolled by the precondition generation.
In the case of infinite iterations, the reachability search space will be infinitely large.
To deal with this search space, we order and prune the space.
Theorem~\ref{thm:presunsat} states a basic property of exploit triples that is used for the purpose of pruning.
Section~\ref{sec:lithmus} describes how the space is ordered to manage large search spaces.

\begin{theorem}[Preservation of unsatisfiability]
For any program $p$ and conditions $P$ and $Q$ such that $\langle P\rangle ~p~ \langle Q\rangle$,
\begin{center}
  $(\forall \State' \cdot Q(\State') \implies \false) \implies (\forall \State \cdot P(\State)\implies \false)$
\end{center}
\label{thm:presunsat}
\end{theorem}

The above can directly be concluded from the definition of a reachability triple, as given in Section~\ref{sec:logic}.
Once an unsatisfiable condition is generated, the precondition generation can be halted, and the condition discarded.

We validate our precondition generation function by proving it is both sound and complete.
Theorems~\ref{thm:sound} and \ref{thm:comp} define these respective properties.

\begin{theorem}[Soundness of precondition generation]
Functions $\tau_P$, $\tau_B$ and $\tau_S$ are sound.
\label{thm:sound}
\end{theorem}

\begin{theorem}[Completeness of precondition generation]
\quad\vspace{-1.3em}\\
  \begin{center}
  $\inferrule*{$\mbox{termination}(p,P)$ \\ $\mbox{no\_indirections}(p)$}{\langle P\rangle ~p~ \langle Q\rangle \implies \exists P' . \preP{p}{Q} \land (P \implies P')}$
  \end{center}
  % \[
  %   \langle P\rangle ~p~ \langle Q\rangle \implies \exists P' . \preP{p}{Q} \land (P \implies P')
  % \]
  \label{thm:comp}
\end{theorem}
% \begin{prooftree}
% \AxiomC{$\mbox{termination}(p,P)$}
% \AxiomC{$\mbox{no\_indirections}(p)$}
% \BinaryInfC{$\ldots \implies \ldots$}
% \end{prooftree}

Having both soundness and completeness means that the reachability space defines all and only valid preconditions for a certain program and postcondition.

Both theorems, including 1.) the syntax and semantics of \jump, 2.) the syntax and semantics of the predicates, and 3.) the functions $\tau$ have been formally proven correct in the Isabelle/HOL theorem prover.
The proof, including a small example of exploit generation within Isabelle/HOL, constitutes roughly 1000 lines of code. Proof scripts are publicly available~\footnote{https://github.com/niconaus/reachability-logic/tree/main/isabelle}.
To prove completeness, Theorem~\ref{thm:comp} imposes two restrictions.
One, we require execution of a program $p$ under a state described by $P$ to terminate.
If a program does not terminate, it is impossible to construct a P' for this program, and therefore completeness does not hold.
Two, we show the theorem holds for programs without indirect jumps.
%e had to include this restriction to complete the proof.
In practice however, this premise has little to no impact.
Every \jump\ program containing indirect jumps, can be manually converted to one with only direct jumps.
Given that $P$ is a precondition for program $p$ and postcondition $Q$, the precondition generation will generate a $P'$ that is non-strictly weaker than $P$.
An equivalent of Theorem~\ref{thm:comp} also holds for $\tau_S$ and $\tau_B$.

\section{Litmus tests}
\label{sec:lithmus}

This section presents several litmus tests that demonstrate the functionality of Reachability Logic.
All of the examples have been tested in our prototype implementation in Haskell, and are available online~\footnote{https://github.com/niconaus/reachability-logic}.
%The implementation is available online, containing the examples shown below. % REMOVED FOR DOUBLE BLIND REVIEW \footnote{https://github.com/niconaus/exploit-logic/tree/master/Haskell}.

The prototype implements the $\pre$ functions similar to how they are presented above.
Some changes were made to make the system more user friendly.
The $\pre$ functions are defined as non-deterministic functions, building up a tree as a search space.
Branches at the same level originate from a conditional, and deeper branches indicate a deeper level of jumps.
On top of that, a basic simplification step is applied to the generated predicates, to make them more readable.

The reachability search space can be infinitely large.
This is why the implementation builds up the search space as a tree structure.
This orders the search space, making it feasible to search the infinite space in a structured way.
Although some rudimentary ordering is done, efficiently searching the reachability space is explicitly left as future work.

\subsubsection{Memory: Double store}

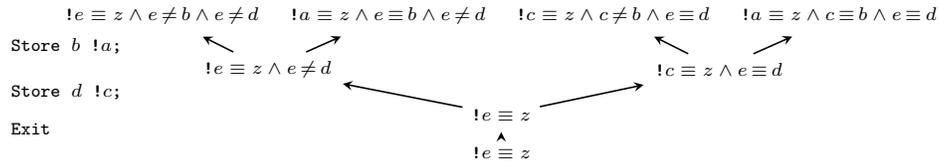
\begin{figure}[t]
  \advance\leftskip-.3cm
  \tikzstyle{sdrive} = [->]

\begin{tikzpicture}[
            > = stealth, % arrow head style
            shorten > = 1pt, % don't touch arrow head to node
            auto,
            node distance = 1cm, % distance between nodes
            semithick, % line style
            font=\scriptsize
        ]

        \node (c1) [rotate=0,align=left,text width=2cm] {$\Store\ b\ \Deref a\Seq$};
        \node (s1) [right of=c1,yshift=.4cm,xshift=.0cm,rotate=0] {\begin{tabular}{c}$\Deref e \equiv z\land e \Separate b\land e \Separate d$\end{tabular}};
        \node (ppp2) [right of=s1,xshift=2cm,yshift=0cm,rotate=0] {\begin{tabular}{c}$\Deref a \equiv z\land e \Aligned b\land e \Separate d$\end{tabular}};
        \node (ppp3) [right of=ppp2,xshift=2cm,rotate=0] {\begin{tabular}{c}$\Deref c \equiv z\land c \Separate b\land e \Aligned d$\end{tabular}};
        \node (ppp4) [right of=ppp3,xshift=2cm,rotate=0] {\begin{tabular}{c}$\Deref a \equiv z\land c \Aligned b\land e \Aligned d$\end{tabular}};

        \node (ck) [below of=c1,xshift=0cm,yshift=.4cm,align=left,text width=2cm] {$\Store\ d\ \Deref c\Seq$};
        \node (sk) [below right of=s1,yshift=0.0cm,xshift=0.7cm] {$\Deref e \equiv z \land e \Separate d$};
        \node (lk) [below right of=ppp3,yshift=-.0cm,xshift=0.7cm] {$\Deref c \equiv z \land e \Aligned d$};

        \node (s5) [below of=ck,yshift=.5cm,align=left,text width=2cm] {$\Exit$};
        \node (sn) [below right of=sk,yshift=.1cm,xshift=2.4cm] {$\Deref e \equiv z$};

        \node (q) [below of=sn,yshift=.5cm] {$\Deref e \equiv z$};

        \draw[sdrive] (sk) -- (s1);
        \draw[sdrive] (sk) -- (ppp2);
        \draw[sdrive] (lk) -- (ppp3);
        \draw[sdrive] (lk) -- (ppp4);

        \draw[sdrive] (sn) -- (sk);
        \draw[sdrive] (sn) -- (lk);

        \draw[sdrive] (q) -- (sn);

    \end{tikzpicture}
    \caption{Precondition generation for double write example}
%    \Description[Precondition generation for double write example]{Precondition generation for double write example}
      \label{fig:doubleStore}
      \vspace{-.5cm}
\end{figure}

Our first litmus test demonstrates how Reachability Logic deals with symbolic memory access.
Figure~\ref{fig:doubleStore} lists the program on the left.
The program stores the content of memory position $a$ in memory position $b$, and then stores the contents of position $c$ in position $d$.
On the right, a schematic representation of the precondition generation is given for postcondition $\Deref e \equiv z$.
Precondition generation works from back to front, so we start at the bottom.
The $\Exit$ command leaves the condition unaltered.
We then arrive at a $\Store$ statement.
Here, we split into two cases, one for the case where memory location $e$ and $d$ are separate, one for the aliasing case.
The second $\Store$ splits the conditions up once more, for the separate and aligned cases.
At the top of the tree, we find the final preconditions.

% \begin{align*}
%   &\Deref e \equiv z \land \Separate e \ d\\
%   &t_2 \equiv z \land \Aligned e\ d\\
% \end{align*}
%
% \begin{align*}
%   &\Deref e \equiv z \land \Separate e \ d\\
%   &\Deref c \equiv z \land \Aligned e\ d\\
% \end{align*}
%
% \begin{align*}
%   &\Deref e \equiv z \land \Separate e\ b\land \Separate e \ d\\
%   &t_1 \equiv z \land \Aligned e\ b\land \Separate e \ d\\
%   &\Deref c \equiv z \land \Separate c\ b\land \Aligned e \ d\\
%   &t_1 a \equiv z \land \Aligned c\ b\land \Aligned e \ d\\
% \end{align*}
%
% \begin{align*}
%   &\Deref e \equiv z \land \Separate e\ b\land \Separate e \ d\\
%   &\Deref a \equiv z \land \Aligned e\ b\land \Separate e \ d\\
%   &\Deref c \equiv z \land \Separate c\ b\land \Aligned e \ d\\
%   &\Deref a \equiv z \land \Aligned c\ b\land \Aligned e \ d\\
% \end{align*}

\subsubsection{Infinite exploit space: Long division}

\begin{figure*}[t]
  \tikzstyle{sdrive} = [->]
  \tikzstyle{unsat} = [dashed,-|]

\begin{tikzpicture}[
            > = stealth, % arrow head style
            shorten > = 1pt, % don't touch arrow head to node
            auto,
            node distance = 1cm, % distance between nodes
            semithick, % line style
            font=\scriptsize
        ]

        % j = 1
        \node (c1) {\begin{tabular}{l}$\#0$\\$ \Jump\ (x < y)\ \#3\ \#1$\end{tabular}};
        \node (s1) [right of=c1,yshift=1cm,xshift=.35cm] {\begin{tabular}{l}$x\geq y\land x< y$\end{tabular}};
        \node (ppp2) [right of=c1,yshift=1.5cm,xshift=3.1cm] {};
        \node (ppp3) [right of=ppp2,xshift=1.6cm] { };
        \node (ppp4) [right of=ppp3,xshift=0cm,yshift=-.6cm] {\begin{tabular}{l}$(x-3y)\geq y\land \neg ((x-3y) >y)$\\$\land ((x-2y)>y)\land (x-y)\geq y \land \neg (x<y)$\end{tabular}};

        \node (cc1)[below of=c1,yshift=-.25cm] {\begin{tabular}{l}$\#1$\\$\Assign\ i\ 1\Seq$\\$\Assign\ x\ (x-y)\Seq$\\$ \Jump\ (x\geq y)\ \#2\ \#3$\end{tabular}};
        \node (cs1) [right of=cc1,yshift=.75cm,xshift=1cm] {};
        \node (cppp2) [right of=cs1,xshift=-.2cm] {\begin{tabular}{l}$(x-y)\geq y $\\$\land \neg ((x-y) \geq y)$\end{tabular}};
        \node (cppp3) [right of=cppp2,xshift=1.3cm] {\begin{tabular}{l}$(x-2y)\geq y $\\$\land \neg ((x-2y) >y)$\\$\land ((x-y)\geq y)$\end{tabular}};
        \node (cppp4) [right of=cppp3,xshift=2.2cm] {\begin{tabular}{l}$(x-3y)\geq y \land \neg ((x-3y) >y)$\\$\land ((x-2y)>y)\land (x-y)\geq y $\end{tabular}};
        \node (cc) [below of=cc1,yshift=-.5cm] {\begin{tabular}{l}$\#2$\\$\Assign\ x\ (x-y)\Seq$\\$ \Assign\ i\ (i+1) \Seq$\\$ \Jump (x > y)\ \#2\ \#3$\end{tabular}};
        \node (pp2) [below of=cppp2, yshift=-1cm] {};
        \node (pp3) [below of=cppp3, yshift=-.7cm] {\begin{tabular}{l}$(x-y)\geq y $\\$\land \neg ((x-y) >y)$\end{tabular}};
        \node (pp4) [below of=cppp4, yshift=-.7cm] {\begin{tabular}{l}$(x-2y)\geq y $\\$\land \neg ((x-2y) >y)$\\$\land ((x-y)>y)$\end{tabular}};
        \node (pp5) [right of=pp4,xshift=1cm] {$\cdots$};

        % j = k
        \node (ck) [below of=cc,yshift=-.2cm,xshift=-.9cm] {\begin{tabular}{l}$\#3$\\$\Exit$\end{tabular}};
        \node (sk) [right of=ck,yshift=.3cm,xshift=1.3cm] {$x\geq y$};
        \node (sk1) [right of=ck,yshift=-.25cm,xshift=1.3cm] {$x\geq y$};

        \node (bang1) [below of=s1,yshift=0.5cm,xshift=.4cm] {$\lightning$};
        \node (bang2) [below of=cppp2,yshift=0.3cm,xshift=.4cm] {$\lightning$};
        \node (bang3) [below of=cppp3,yshift=0.2cm,xshift=.4cm] {$\lightning$};

        \draw[unsat] (pp3) -- (cppp3);
        \draw[sdrive] (pp4) -- (cppp4);
        \draw[sdrive] (cppp4) -- (ppp4);

        \draw[unsat] (sk) -- (cppp2);

        \draw[unsat] (sk) -- (s1);
        \draw[sdrive] (sk) -- (pp3);
        \draw[sdrive] (sk1) -- (sk);

        \draw [->] (pp3) to [out=-90,in=-135] (pp4);
        \draw [->] (pp4) to [out=-90,in=-135] (pp5);

    \end{tikzpicture}
    \caption{Precondition generation for long division example. A dashed arrow leads to an unsatisfiable precondition.}
%    \Description[Precondition generation for long division example]{Precondition generation for long division example}
      \label{fig:longdivision}
      \vspace{-.5cm}
\end{figure*}
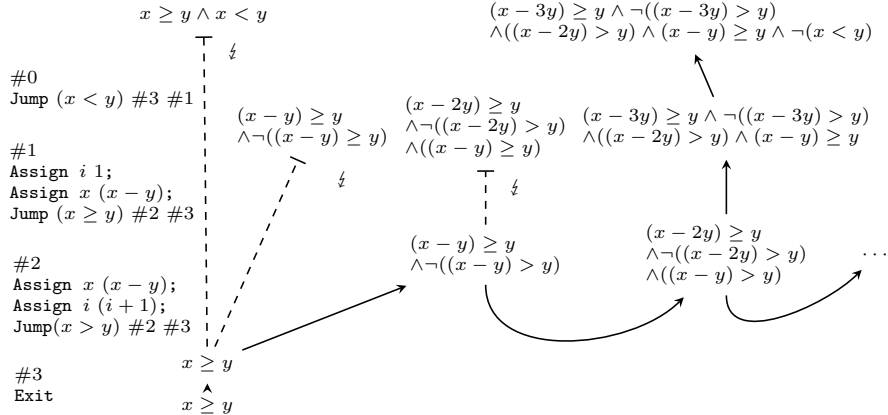

Our next litmus test demonstrates conditional jumps, loops, infinite reachability space and postcondition pruning.
Figure~\ref{fig:longdivision} lists the program blocks on the left.
The blocks are labeled $\#0$ though $\#3$, with block $\#0$ is the entry point.
Variables $x$ and $y$ signify the input.
The program divides $x$ by $y$, by means of long division.
If $x$ is larger than $y$, the divisor without remainder is returned in variable $i$.
The variable $x$ is updated, and after execution holds the remainder from division.

In this case, we want to derive that a state is reachable which clearly should not be, to show that there is a bug in the program.
The program behaves incorrectly when after execution, the remainder stored in $x$ is equal to or larger than the divisor $y$.
We use this, $x\geq y$, as our exploit postcondition.

The right of Figure~\ref{fig:longdivision} represents precondition generation.
We start back to front.
$\Exit$ does not alter the postcondition, so we just copy it.
Then, we either execute block 0, 1 or 2, depending on what condition holds.
If we came directly from block 0, then $x<y$ must hold, so our precondition is $x\geq y\land x<y$, which is false, indicated by the lightning bolt.
If we came from block 1, then $\neg(x\geq y)$ must have held.
Block 1 updates $x$ with $x-y$, leading to the precondition $(x-y)\geq y \land \neg((x-y)\geq y)$.
Note that this precondition is unsatisfiable.
By Theorem~\ref{thm:presunsat} we know that we can discard it.

The last block to look at, is block 2.
To arrive here, we must have had that $\neg (x>y)$.
The body of block 2 updates $x$, and we end up with $(x-y)\geq y \land \neg ((x-y)>y)$.
Here, we see the loop unfolding at work.
We have executed the loop body once, and the $\pre$ function generates two alternatives.
We exit the loop, indicated by the arrow pointing up, or we run another iteration, indicated by the arrow pointing right.

Ending the loop at this point again leads to a precondition that is unsatisfiable, and we can prune it.
Running the loop a second time, and then exiting leads to a precondition that is satisfiable, and completing the calculation, leads us to the first viable precondition for the postcondition $x\geq y$.

The precondition function $\pre$ does not stop at this point.
It will continue to unroll the loop an infinite amount of times, making the exploit space infinitely large.
%The loop could be unrolled a third time, a fourth, and in fact an infinite amount of times.
%Therefore the exploit space for this example is infinitely large.
By ordering the space as shown in this example, we can deal with infinity.
% \subsection{Nondeterminism: Dice}
%
% The example in this section demonstrates nondeterminism.
% Figure~\ref{fig:dice} lists a program consisting of a single statement, representing the trow of a die.
% To the variable $die$, a random integer value from one to six is assigned.
%
% \begin{figure}
%   \tikzstyle{sdrive} = [->]
%
% \begin{tikzpicture}[
%             > = stealth, % arrow head style
%             shorten > = 1pt, % don't touch arrow head to node
%             auto,
%             node distance = 1cm, % distance between nodes
%             semithick, % line style
%         ]
%
%         % j = 1
%         \node (c1) {$\Assign die\ \From~ \{ mv \mid 1 \leq mv \leq 6 \}$};
%         \node (s1) [right of=c1,yshift=.5cm,xshift=2cm] {$\exists i . i = 6 \land 1 \leq i\leq 6$};
%
%         %
%         \node (ss) [below of=s1] {\begin{tabular}{l}$die = 6$\end{tabular}};
%
%         \draw[sdrive] (ss) -- (s1);
%
%     \end{tikzpicture}
%     \caption{Precondition generation for dice example}
%       \label{fig:dice}
% \end{figure}
%
% Imagine we are interested in throwing a six.
% We can then formulate the postcondition $die = 6$.
% Running the precondition generation, will then give us in what conditions this will occur.
% Looking on the left side, we see that the precondition that must hold is $\exists i . i = 6 \land 1 \leq i\leq 6$.
% Since this is always the case, we can simplify this to $true$.

\subsubsection{Nondeterministic loop}

\begin{figure}[t]
  \tikzstyle{sdrive} = [->]

\begin{tikzpicture}[
            > = stealth, % arrow head style
            shorten > = 1pt, % don't touch arrow head to node
            auto,
            node distance = 1cm, % distance between nodes
            semithick, % line style
            font=\scriptsize
        ]

        % j = 1
        \node (c1) {\begin{tabular}{l}$\#0:
                                      \NDAssign\ d\ \From~ 1 < d < n;
                                       \Jump\ (n\% d \equiv 0)\ \#1\ \#0$\end{tabular}};
        \node (s1) [below of=c1,xshift=-2cm,yshift=.5cm] {\begin{tabular}{l}$\exists i.(n\% i \equiv 0)\land (1< i)\land(i<n)$\end{tabular}};
        \node (ppp2) [below of=s1,yshift=.25cm,xshift=2cm] {\begin{tabular}{l}$\exists i.(n\% i \equiv 0)\land (1< i)\land(i<n)
        \land\exists j. \neg((n \% j) \equiv 0)\land (1 < j) \land (j < n)$\end{tabular}};
        \node (ppp3) [below of=ppp2,yshift=.25cm] {$\cdots$};

        \node (ck) [right of=c1,xshift=4cm] {\begin{tabular}{l}$\#1:\Exit$\end{tabular}};
        \node (sk) [below of=ck,xshift=-.8cm,yshift=.5cm] {$\true$};
        \node (sk1) [below of=ck,xshift=.9cm,yshift=.5cm] {$\true$};

        \draw[sdrive] (sk) -- (s1);

        \draw[sdrive] (sk1) -- (sk);

        \draw [->] (s1) to [out=-5,in=15] (ppp2);
        \draw [->] (ppp2) to [out=-15,in=15] (ppp3);

    \end{tikzpicture}\vspace{-.4cm}
    \caption{Precondition generation for nondeterministic loop example}
%    \Description[Precondition generation for nondeterministic loop example]{Precondition generation for nondeterministic loop example}
      \label{fig:ndloop}
      \vspace{-.5cm}
\end{figure}
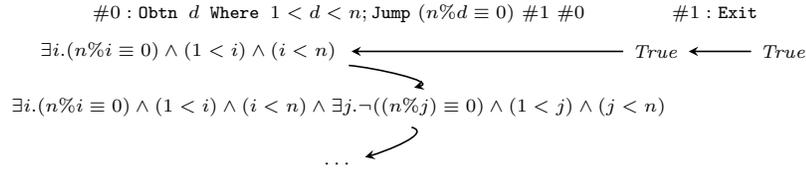

A defining feature of both \jump\ and Reachability Logic, is the ability to deal with nondeterminism.
The program listed in Figure~\ref{fig:ndloop} is a very small example of such a program.
It assigns a random integer value to the variable $d$, between 1 and some value $n$.
Then, if $n$ modulo $d$ is zero, the program jumps to $\Exit$, otherwise it repeats itself.

For this example, we are simply interested in termination.
So as a postcondition, we pick $\true$.
The $\Exit$ block does not change the condition.
Executing the loop once, means that $n\%d \equiv 0$ must hold.
The nondeterministic assignment updates $d$ with any value between 1 and $n$.
Translated to precondition, this means that there exists an $i$ such that $(n\% i \equiv 0)\land (1< i)\land(i<n)$.

This condition holds if and only if $n$ is not prime.
What makes this example interesting is the contrast with other program logics like Hoare logic.
With Hoare Logic, we are only able to prove that if the program terminates, $n$ was not prime.
Reachability Logic however allows us to show that if $n$ is not prime, then the program can terminate, which is a much more powerful statement.

\subsubsection{Indirect Jumps}

\begin{figure}[t]
   \advance\leftskip-.3cm
  \tikzstyle{sdrive} = [->]
  \tikzstyle{unsat} = [dashed,-|]

\begin{tikzpicture}[
            > = stealth, % arrow head style
            shorten > = 1pt, % don't touch arrow head to node
            auto,
            node distance = 1cm, % distance between nodes
            semithick, % line style
            font=\scriptsize
        ]

        % j = 1
        \node (s0) {\begin{tabular}{l}$\#0: \Assign\ y \ 0;$\\
                                      $\phantom{\#0: } \Jump$\\
                                      \quad$\phantom{\#0: } (x < 0\ ||\ x > 3 )$\\
                                      \quad$\phantom{\#0: }   \#9\ \#1$\end{tabular}};
            \node(pre0) [right of=s0,yshift=1.5cm,xshift=.0cm,rotate=25] {\begin{tabular}{l}$0>0 $\\$\land(x < 0 \lor  x >3)$\end{tabular}};
            \node(pre02)[right of=pre0,xshift=1.0cm,rotate=25] {\begin{tabular}{l}$0>0 \land x \equiv 9 $\\$\land x \geq 0 \land  x \leq 3$\end{tabular}};
            \node(pre03)[right of=pre02,xshift=1.2cm,rotate=25] {\begin{tabular}{l}$0>0 \land x \equiv 1 $\\$\land x \geq 0 \land  x \leq 3$\end{tabular}};
            \node(pre00)[right of=pre03,xshift=1.2cm,rotate=25] {\begin{tabular}{l}$5>0 \land x \equiv 0 $\\$\land x \geq 0 \land  x \leq 3$\end{tabular}};
            \node(pre01)[right of=pre00,xshift=1.2cm,rotate=25] {\begin{tabular}{l}$5>0 \land x \equiv 2 $\\$\land x \geq 0 \land  x \leq 3$\end{tabular}};

        \node (s1) [below of=s0,yshift=-.2cm,xshift=-.40cm] {\begin{tabular}{l}$\#1: \IJump\ x$\end{tabular}};
            \node(pre1) [below of=pre00,yshift=-1.2cm,rotate=0] {$5>0 \land x \equiv 2$};
            \node(pre11) [right of=pre1,yshift=-.4cm,xshift=-.15cm,rotate=0] {$\land x \equiv 1$};
            \node(pre13) [below of=pre01,yshift=-1.2cm,rotate=0] {$5>0 \land x \equiv 4$};
            \node(pre14) [right of=pre13,yshift=-.4cm,xshift=-.15cm,rotate=0] {$\land x \equiv 1$};
            \node(pre15) [below of=pre02,yshift=-1.2cm,rotate=0] {$y>0 \land x \equiv 9$};
            \node(pre16) [right of=pre15,yshift=-.4cm,xshift=-.15cm,rotate=0] {$\land x \equiv 1$};
            \node(pre17) [below of=pre03,yshift=-1.2cm,rotate=0] {$y>0 \land x \equiv 3$};
            \node(pre18) [right of=pre17,yshift=-.4cm,xshift=-.15cm,rotate=0] {$\land x \equiv 1$};

        \node (s2) [below of=s1,yshift=-.0cm,xshift=.4cm] {\begin{tabular}{l}$\#2: \Jump\ true\ \#4\ \#4$\end{tabular}};
            \node (pre2) [below of=pre1,yshift=-.0cm] {$5>0$};

        \node (s3) [below of=s2,yshift=.4cm] {\begin{tabular}{l}$\#3: \Jump\ true\ \#9\ \#9$\end{tabular}};
            \node (pre3) [below of=pre17,yshift=-.7cm] {$y>0$};

        \node (s4) [below of=s3,yshift=.3cm] {\begin{tabular}{l}$\#4: \Assign\ y\ 5;$\\
                                                           $\phantom{\#2: }\Jump\ true\ \#9\ \#9$\end{tabular}};
            \node (pre4) [right of=s4,yshift=.5cm,xshift=8cm] {$5>0$};

        \node (s5) [below of=s4,yshift=.3cm,xshift=-.8cm] {\begin{tabular}{l}$\#9: \Exit$\end{tabular}};
          \node (pre5) [below of=pre4,yshift=-.0cm,xshift=-6cm] {$y>0$};
          \node (post) [below of=pre5,yshift=.3cm] {$y > 0$};

        \draw[sdrive] (post) -- (pre5);
        \draw[unsat] (pre5) -- (pre0);
        \draw[sdrive] (pre5) -- (pre15);
        \draw[sdrive] (pre5) -- (pre3);
        \draw[sdrive] (pre5) -- (pre4);
        \draw[sdrive] (pre4) -- (pre2);
        \draw[sdrive] (pre4) -- (pre13);
        \draw[sdrive] (pre3) -- (pre17);
        \draw[sdrive] (pre2) -- (pre1);
        \draw[sdrive] (pre1) -- (pre00);
        \draw[sdrive] (pre13) -- (pre01);
        \draw[unsat] (pre15) -- (pre02);
        \draw[unsat] (pre17) -- (pre03);

        \draw [unsat] (pre1) to [out=-100,in=-90] (pre11);
        \draw [unsat] (pre13) to [out=-100,in=-90] (pre14);
        \draw [unsat] (pre15) to [out=-100,in=-90] (pre16);
        \draw [unsat] (pre17) to [out=-100,in=-90] (pre18);

        \node (bang1) [below of=pre11,yshift=.5cm,xshift=.2cm] {$\lightning$};
        \node (bang2) [below of=pre14,yshift=.5cm,xshift=.2cm] {$\lightning$};
        \node (bang3) [below of=pre16,yshift=.5cm,xshift=.2cm] {$\lightning$};
        \node (bang4) [below of=pre18,yshift=.5cm,xshift=.2cm] {$\lightning$};
        \node (bang6) [below of=pre02,yshift=-.2cm,xshift=.4cm] {$\lightning$};
        \node (bang7) [below of=pre03,yshift=-.2cm,xshift=.4cm] {$\lightning$};
        \node (bang8) [below of=pre0,yshift=-.2cm,xshift=.8cm] {$\lightning$};

    \end{tikzpicture}
    \caption{Precondition generation for indirect jump example.}
%    \Description[Precondition generation for indirect jump example]{Precondition generation for indirect jump example}
      \label{fig:indirect}
      \vspace{-.5cm}
\end{figure}
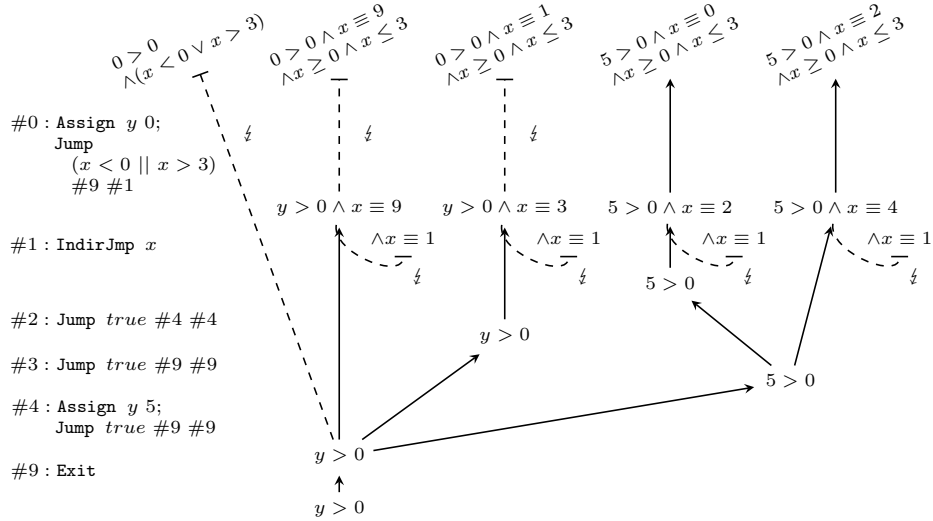

Our last litmus test demonstrates how Reachability Logic deals with indirect jumps.
Switch-statements consisting of many cases are often compiled into jump tables. These are typically combined
with a guard for values not handled by the jump table.
Figure~\ref{fig:indirect} shows a model of this.

Execution starts at block 0.
Here, $y$ is set to 0, and the conditional jump checks if x is smaller than 0 or larger than 3.
If so, we jump to exit.
If not, we jump to block 1, which is the start of our guard.
The indirect jump jumps to the block label stored in $x$.
Blocks 2, 3, and 4 signify the guard options.

The goal of this example is to demonstrate how an indirect jump works with our precondition generation function.
As a postcondition, we select $y>0$.

Starting at block 9, we again work our way up the execution back to front.
Block 9 itself has no effect on the postcondition, so it is copied.
We can reach block 9 from four different locations in the code, namely block 0, 1, 3 and 4.

Block 0 assigns 0 to $y$, which leads to a contradiction in our precondition, namely $0>0$.
Block 1 is the indirect jump.
We can only have reached block 9 from here if $x$ was equal to 9.
In turn, we could have reached block 1 from itself, or 0.
If we indeed came from itself, we must have made an indirect jump to 1, requiring x to be equal to 1 as well, which is a contradiction.
If we came from 0, we have the same problem as before, $y$ is set to 0 and so $0>0$ must hold.

Block 3 always jumps to 9, so it has no effect on the condition.
The only way to reach block 3 is from block 1, the indirect jump.
This requires x to be equal to 3.
We can reach block 2 from itself and 0.
Both lead to a contradiction; if we came from 1, then $x \equiv 1$, which is false, if we came from 0, $0>0$ must hold.

Block 4 is our last hope to find a satisfiable precondition.
The block itself assigns 5 to $y$, leading to the precondition $5>0$.
Block 4 can be reached from block 2 or 1.
Block 2 is a simple jump, so we copy the condition.
Block 2 is only reachable from block 1, which adds $x \equiv 2$ to the condition.
We can only reach 1 from itself or 0.
If we came from itself, we get a contradiction immediately.
If we came from 0, we know that the jump condition was false, so we add its negation to the condition.
This leads to the first satisfiable precondition.

If we came directly from block 1, we know that $x \equiv 4$ should hold.
Analogous to the case described directly above, we obtain that the only feasible route is via block 0, which again leads to a satisfiable precondition.

What is interesting about this example, is the fact that although we have an indirect jump, the number of paths to explore stays rather limited.
Potentially, an indirect jump can jump to any address, but in practice, these addresses are limited by the conditions that must hold.

% !TEX root=../main.tex

\section{Case Studies}\label{sec:casestudies}

We presents results from applying Reachability Logic to two realistic examples. These case studies were performed using
the prototype mentioned in Section~\ref{sec:lithmus}.

\subsubsection*{Faulty Partitioning for Quicksort}
The core of any quicksort algorithm is the partitioning algorithm. One well-known partitioning algorithm is the one invented
by Tony Hoare~\cite{hoare1961algorithm64} which selects a \emph{pivot} element and then transforms an input data set into two smaller sets,
depending on relative ordering of elements in the data set to the pivot.
This scheme seems superficially very simple, but it is very easy to get wrong.
For instance, the following algorithm has a superficially plausible variant of this partitioning scheme, which is ``nearly correct''.
{\footnotesize
\begin{verbatim}
void quicksort(int a[], size_t N) {
  if(N <= 1) return;
  int pivot = a[rand()%N];
  int i = 0, j = N-1;
  while(i <= j) {
    while(i < j  && a[i] <= pivot) i++;
    while(i <= j && a[j] >= pivot) j--;
    swap(&a[i++], &a[j--]);
  }
  quicksort(a, j+1);
  quicksort(a+i, N-i);}
\end{verbatim}
}
%While \jump\ does not support recursion directly,
The partitioning scheme can be
translated into a \jump\ program relatively easily; selection of the pivot can be modeled using
a non-deterministic assign.

We are interested in detecting out-of-bounds memory access.
We add bounds checks to the program, and state triggering one of them as our postcondition.
Running the resultant program through our implementation for an array of size 3 will then
generate an exploit-precondition: the program can go out of bounds if the following condition holds:
$$
\exists i. 0 \le i \le 3 \land a[i] \le a[0] \land a[i] \le a[1] \land a[i] \le a[2] \land a[0] > a[i]
$$
Informally, this conditions says that \texttt{a[0]} is not the minimal element of the array. The reason for this is that
if the minimal element is chosen as a pivot, and \texttt{a[0]} is not equal to it, the first inner loop will simply fall through,
and after the second loop, $i$ will become $-1$, pointing outside the array before the swap occurs.
A fix for this would be make the swap conditional, replacing it with:
\begin{verbatim}
    if(i <= j) swap(&a[i++], &a[j--]);
\end{verbatim}

This will in fact prevent any out-of-bound memory access. However, another way any version of quicksort can fail dramatically is
when the recursive calls
are performed with incorrect parameters. For example if $i = 0$ or $j=N$ at the end of the partitioning scheme, we will end up in a
infinite recursive loop. If we specify this as a postcondition of the partitioning scheme, we find that the same
preconditions
are generated as before.

The functional correctness of the partitioning scheme can also be examined---that is, is it actually the case that all the elements moved towards the the left-hand side of the array are less-or-equal to the pivot, and that the elements to the right are greater-or-equal than the pivot? To examine this, we can specify as an exploit condition that the input to the first recursive invocation of \texttt{quicksort} contains an element greater than the pivot; this finds no satisfiable conditions (as it is not true).
However, specifying this for input sent to the second invocation of \texttt{quicksort} instead,
our prototype will essentially start generating counter-examples.
%For example, if the last element is the maximum of the array, and also selected as the pivot, and the middle element is strictly less than the pivot, the partitioning fails.
For example, if the first element is the pivot, and strictly less than the middle element but strictly higher than the third element, partitioning fails.

\subsubsection*{Karatsuba}
Several assembly routines for multiplying multi-precision integers on an 8-bit AVR controller were verified
by Schoolderman \cite{schoold2017asmwhy3}.
However, it was also discovered that some of these routines could compute incorrect results if their arguments aliased
with the memory location intended to store the result. A full verification like this appears to require significant effort;
however, if we are only interested in finding aliasing bugs, RL seems ideally suited to find these.

We focused on the smallest routine exhibiting the problem: the $48\times 48\to 96$-bit multiplication routine as originally
developed by Hutter and Schwabe \cite{hutter2015mul}.
This routine computes a product of two 48-bit integers using Karatsuba's method, splitting its inputs into two 24-bit halves, and performing a
three $24\to48$-bit multiplications with these, combining the results.\footnote{To be more precise, this method uses the fact
that $(2^w X_h+X_l)(2^w Y_h + Y_l) = (1+2^w)(2^w X_hY_h + X_lY_l) - 2^w(X_l-X_h)(Y_l-Y_h)$}
In the process, the lowest 24-bits of the result are known early on and written to memory before the upper half of the
inputs is read, causing an aliasing bug.

To model this in \jump, registers and the carry flags are modeled as \jump\ variables, whereas the memory space is modeled using
\jump\ addresses. Every AVR instruction is modeled by a sequence of \jump\ statements. For example, the instruction \textsc{ADD $a0$, $a1$} can
be expressed by the sequence:
\begin{align*}
	&\Assign\ tmp\ (a0 + a1);\
	\Assign\ a0\ (tmp~\text{mod}~256);\
	\Assign\ carry\ (tmp~/~256)
\end{align*}

Adding the appropriate binary operators to the syntax of Figure~\ref{fig:syntax}, every instruction required for the program (which are only a handful) can be modeled, allowing the entire multiplication routine (consisting of 136 instructions) to be expressed as a \jump\ program. The memory accesses, which operate on three bytes at a time, were modeled as a single memory operations on a three-byte memory region.

As seen in Section~\ref{sec:lithmus}, generated preconditions can be fairly verbose, and we expected that in this case as well.
To remedy this somewhat, we extended the Haskell implementation with constant folding and other
simplifications to more efficiently manage the search space of possible preconditions, and pruning areas of the search space
which can easily be determined to be impossible. In a more production-oriented setting, SMT solving and/or a robust expression simplifier can be used to do this more efficiently than our naive Haskell implementation.

For the precondition, we look at the case $X\cdot Y$ where $X=Y=2^{24}$. Clearly the expected result
should be $X\cdot Y = 2^{48}$, i.e. the 96-bit result should consist of 12 bytes, all of which contain $0$, except for the
seventh byte which should hold $1$. As a postcondition, we therefore specify that this byte does \emph{not} hold $1$.

Running the \jump\ version of the 48-bit Karatsuba code through our analysis resulted in a handful of preconditions. Some
of these simplify to $\false$, as they express impossible aliasing conditions---an SMT solver would be able to discard these
easily. However, 7 preconditions remained which are completely plausible and satisfiable, which fall into three categories:
\begin{itemize}
	\item $X$, $Y$ alias, and their high 24-bits overlap with the low 24-bits of the result
	\item $X$, $Y$ are disjoint, and of them partially overlaps with the result as before
	\item $X$, $Z$ are partially aliased, and one of them partially overlaps with the result
\end{itemize}

Which are exactly the case we would expect: the issue is being caused by either (or both) inputs sharing their high 24-bits
with the low 24-bits of the output location.
Had we not chosen the fixed input values for $X$ and $Y$, this case would
have generated more complex preconditions, however, this case shows that there is an easy instance where these would be
satisfied.

\section{Discussion}\label{sec:discussion}

Key to our approach is the reachability space.
The size and shape of this search space directly influences the effectiveness or even feasibility of Reachability Logic.
State space explosion is a very typical and problematic effect for approaches like symbolic execution and loop unrolling.

How is this different for Reachability Logic?
First of all, the goal of RL is not to say something about all possible execution paths, but instead only one.
This means that instead of proving properties over the entire search space, we just need to find one execution path that is viable.
The nature of the reachability space is such that it only contains paths that are finite in length, but there may be infinitely many of them due to loop unrolling and indirect jumps.
This means that we only need to search though this space in width, not depth, to find one path to the assertion.
On top of that, the reachability space can be ordered in such a way to make searching more efficient.
For the litmus tests and case studies, this is already done to improve performance.

% !TEX root=../main.tex

\section{Related work}\label{sec:related}

Besides the program logics mentioned in Section~\ref{sec:logic}, several other angles have been explored to (automatically) reason over programs, reachability and exploits.

%\subsubsection*{Automatic exploit generation}

%Some research is done on automatically generating exploits.
Brumley et al.~\cite{DBLP:conf/sp/BrumleyPSZ08} generate exploits from patches.
Given a program $P$ and a patched version of the program $P'$, they are able to find the exploit that was patched in $P'$, but to which $P$ is vulnerable.
%This approach can be used to target systems that are out of date.
An obvious drawback of this approach is that a patched version of the program needs to be available, for this approach to work.

Avgerinos et al.~\cite{DBLP:conf/ndss/AvgerinosCHB11} present an automatic exploit generation system (AEG) that only requires the source code of the program to be exploited.
They generate LLVM code from the source and analyze this code using symbolic execution and some safely property to obtain exploits.
% After some preprocessing, they take the code and some safety property, and then look for bugs using symbolic execution.
% The safety property is used to insert checks during symbolic execution.
% To deal with state space explosion, they combine a number of different analysis techniques with search heuristics to detect bugs.
% When a potential bug has been found, it is used to generate the exploit.
% AEG then tests the exploit against a normal binary that has been built with GCC from the original source code.
% They verify their approach by running a number of test programs, which uncovered fourteen programs that contain vulnerabilities.
% Their results are however experimental.
% AEG is in no way formally verified.
Their approach uses heuristics, and the safety property used is not discussed or listed in the paper.
No guarantees are given about the search space, as opposed to our solution, where we formally verify the reachability search space.

%\subsubsection*{Automatic vulnerability detection}

As an alternative to AEG, automatic vulnerability detection is a less strictly defined line of research into finding vulnerabilities.
Russell et al.~\cite{DBLP:conf/icmla/RussellKHLHOEM18} use deep representation learning to do the job.
They use the large set of open-source code available online, labeled with vulnerability information, to train their tool.
Their results are very promising, but as with all machine learning-based solutions, their approach does not come with any guarantees.

%\subsubsection*{Symbolic (backwards) execution}

%Precondition generation, as used in our approach, bears a resemblance with symbolic execution~\cite{boyer1975select,burch1992symbolic,cadar2008klee}.
Symbolic execution runs a program with symbols instead of actual input.
Running the program with these symbolic inputs results in a complete overview of the programs behavior.
%AEG described in the section above relies on symbolic execution to find exploits.
%This is done by inserting assertions in the original code.
Symbolic execution is extensively used for software testing~\cite{boyer1975select,burch1992symbolic,cadar2008klee}.
Cadar and Sen~\cite{DBLP:journals/cacm/CadarS13} provide a great overview of the applications of symbolic execution for this purpose.
The biggest downside of symbolic execution is that it describes the complete program behavior, and therefore quickly becomes infeasible, due to the may paths to be described.
%We are not interested in complete program behaviour, we only want to show that a specific result state can be reached.

%\subsubsection*{Symbolic backward execution}

Symbolic backward execution (SBE) attempts to mitigate the downside of reasoning over all possible paths by looking at a target location in source code.
Charreteur and Gotlieb present a method for generating test input based on SBE for Java bytecode~\cite{DBLP:conf/issre/CharreteurG10}.
Dinges and Agha augment this approach with concrete execution as well~\cite{DBLP:conf/kbse/DingesA14}.
Although there are some similarity between SBE and RL, there are two crucial difference.
First of all, RL is proven to be sound and complete, providing the user with guarantees over the result.
Second, RL is able to deal with non-determinism, modeling external functions, uncertain semantics of some instructions, etc.
Finally, RL only looks at the postcondition, instead of a location in code.
This means that only those statements that affect the postcondition, need to be taken into account, making RL much more feasible than SBE.

%  except for the goal that they have in mind.
% With EL, we want to find one state that will lead to the exploit state.
% SBE on the other hand is interested in finding an initial state that will lead to some point in the source code.
% Looking only at the postcondition has a huge benefit.
% Instead of trying to capture the complete program behavior, only those statements that affect the postcondition need to be taken into account.
% Therefore, the Exploit Logic is much more feasible than SBE.

% Instead, a target location in the source code is selected.
% SBE then backtracks from that point to the entry point of the code.
% The result of the analysis is usually a set of constraints on the machine state, that will lead execution to the desired code location.
%
%
% Starting from the target line of bytecode, the algorithm tries to find a path to the entry point of the bytecode.
% Each program block is associated with constraints that must hold in order to reach it.
% Path feasibility is verified with each code block that is added to the path by verifying the accumulated constraints.
% When the entry point has been reached, their system will attempt to generate concrete values based on the constraints for that path.
% These concrete values form the test input that will lead execution to the desired program point.
%
% They include concrete execution to deal with loops, external methods and undecidable path conditions.

% Weakest precondition generation using backward symbolic execution~\cite{DBLP:conf/pldi/ChandraFS09}

%\subsubsection*{Static bug-finding}

There is a wide variety available of static bug-finding techniques available.
Static analysis has been applied to detect security bugs~\cite{DBLP:conf/www/HuangYHTLK04}, runtime errors~\cite{DBLP:conf/paste/HovemeyerSP05} and even type errors~\cite{DBLP:conf/sas/JensenMT09}.
These analyses are used in practice to detect bugs, for example the FindBug application~\cite{DBLP:journals/software/AyewahHMPP08}, which employs simple static techniques to find bugs in Java code.
Static analyses have to be tailored to a specific type of bugs.
Reachability Logic on the other hand, enables the detection of the reachability of any postcondition.

%\subsubsection*{Test case generation}
%A final relevant line of work is that of test case generation.
Anand et al.~\cite{DBLP:journals/jss/AnandBCCCGHHMOE13} provide a nice overview of the field of test case generation.
They identify five techniques for test case generation.
Symbolic execution, model based testing, combinatorial testing, adaptive random testing and search-based testing.
In general, our approach is a much more formal one than existing test case generation.
We are interested in formally defining the entire reachability space in such a way that it is sound and complete.
Test case generation has a different goal in mind, namely selecting suitable test cases for a certain program.
Many of the techniques mentioned do not guarantee full coverage.
This is something that Reachability Logic does have.

% \subsubsection*{Separation logic}
%
% Separation logic~\cite{DBLP:conf/lics/Reynolds02} is an extension of Hoare logic.
% It allows reasoning about programs that work with shared structures, like the global memory in \jump.
% By making use of virtual separation, reasoning about these kinds of programs is made more scalable.
% Parts of the program under investigation can be separately reasoned over, if it is known which areas of memory they exclusively access.
%
% Like Hoare logic, separation logic reasons about overapproximative behavior.
% This is a crucial difference between Exploit Logic and separation logic; we are only interested in finding a single execution that results in an exploit, instead of describing complete program behavior.
% As mentioned above, separation logic works by separating the memory that is accessed by a piece of code, from the rest of the memory.
% This may require manual effort in the reasoning, since this information must be available to perform separation.
% Exploit Logic and its precondition generation function work completely automatic, as no loop invariants or separation information is required.

% !TEX root=../main.tex

\section{Conclusion}\label{sec:conclusion}

In this paper, we have presented the novel program logic Reachability Logic.
RL is well suited for proving that an assertion is reachable in a nondeterministic program.
Existing program logics such as HL and RHL have a different focus and are unsuitable for this purpose.

We have developed a low-level language called \jump, having jumps instead of explicit control flow.
For this language we presented a precondition generation function, which we have formally proven to be sound and complete.
This precondition function thus generates a sound and complete reachability space.
To validate our approach, we have presented several litmus tests that illustrate the precondition generation function, as well as real world case studies that find bugs.
An implementation of the system has been developed, in which both the litmus tests and case studies have been run.
The entire system is formally proven correct in the Isabelle/HOL theorem prover.

%We have compared our work to the huge body of research into find faults in computer software.
Test case generation research offers a plethora of techniques to show errors in software by means of finding the right test.
While there are many different ways to generate test cases, full coverage is not guaranteed.
Finding vulnerabilities has also been done using machine learning, but similarly this method too does not come with any guarantees.
Symbolic execution comes closer, since it is able to prove properties over all paths in the program, but state space explosion prevents it from being applied to large code bases.
Some initial work has been done on automatic exploit generation.
Compared to existing work, we take a \emph{formal} approach, so that we are able to give the user the guarantee that all preconditions are sound.

\subsubsection{Acknowledgements}

This work is supported by the Defense Advanced Research Projects Agency (DARPA) under Agreement No. HR00112090028 and contract N6600121C4028 and US Office of Naval Research (ONR) under grant N00014-17-1-2297.

% \subsubsection{Acknowledgements} Please place your acknowledgments at
% the end of the paper, preceded by an unnumbered run-in heading (i.e.
% 3rd-level heading).

\bibliographystyle{splncs04}
\bibliography{main}

\end{document}